\documentclass[%
 reprint,
nofootinbib,
 bibnotes,
 amsmath,amssymb,
 aps,
 pra,
]{revtex4-2}

\usepackage{graphicx}
\usepackage{dcolumn}
\usepackage{bm}

\usepackage{subcaption}
\usepackage{upgreek}
\usepackage{siunitx}
\usepackage[symbol]{footmisc}

\begin{document}

\preprint{APS/123-QED}

\title{Room-temperature emission of muonium from aerogel and zeolite targets}

\renewcommand*{\thefootnote}{\fnsymbol{footnote}}

\author{A.~Antognini\textsuperscript{1,2}}
\author{P.~Crivelli\textsuperscript{1}}
\author{L.~Gerchow\textsuperscript{1}}
\author{T.~D.~Hume\textsuperscript{1}} 
\author{K.~Kirch\textsuperscript{1,2}}
\author{A.~Knecht\textsuperscript{2$\dagger$ }\footnotemark[2]}
\author{J.~Nuber\textsuperscript{1,2$\ast$ }\footnotemark[1]}
\author{A.~Papa\textsuperscript{2,3}}
\author{N.~Ritjoho\textsuperscript{1,2}}
\author{M.~Sakurai\textsuperscript{1}}
\author{A.~Soter\textsuperscript{1,2$\ddagger$}\footnotemark[3]}
\author{D.~Taqqu\textsuperscript{1}}
\author{S.~M.~Vogiatzi\textsuperscript{1,2}}
\author{J.~Zhang\textsuperscript{1}}
\author{L.~Ziegler\textsuperscript{2}}

\affiliation{1 Institute for Particle Physics and Astrophysics, ETH Zurich, 8093 Zurich, Switzerland}%
\affiliation{2 Paul Scherrer Institute, 5232 Villigen–PSI, Switzerland}%
\affiliation{3 University of Pisa and INFN, L. Bruno Pontecorvo, Edificio C, 56127 Pisa PI, Italy}

\date{\today}

\begin{abstract}
A low-emittance, high-intensity atomic beam of muonium ($\mathrm{M}=\upmu^+ + \mathrm{e}^-$) using superfluid helium as muon-to-muonium converter is being developed at the Paul Scherrer Institute (PSI). This beam could advance laser spectroscopy of muonium and allow the first atomic interferometry experiments for the direct observation of the M gravitational interaction. In this paper, we describe the development of compact detection schemes which resulted in the background-suppressed observation of atomic muonium in vacuum, and can be adapted for cryogenic measurements. Using these setups, we compared the emission characteristics of various muonium production targets using low momentum ($p_{\mu} = 11$-$13~$MeV/c) muons, and observed muonium emission from zeolite targets into vacuum for the first time. For a specific laser-ablated aerogel target, we determined a muon-to-vacuum-muonium conversion efficiency of $7.23 \pm 0.05 \text{(stat)} ^{+1.06}_{-0.76}\text{(sys)}\SI{}{\percent}$, assuming thermal emission of muonium. Moreover, we investigated muonium-helium collisions and from it we determined an upper temperature limit of 0.3~K for the superfluid helium converter. 
\end{abstract}

\maketitle


\section{Introduction}
\footnotetext[1]{jonas.nuber@psi.ch}
\footnotetext[2]{a.knecht@psi.ch}
\footnotetext[3]{asoter@ethz.ch}
\renewcommand*{\thefootnote}{\arabic{footnote}}
\setcounter{footnote}{0}

%
Muonium (M) is a two-body exotic atom consisting of a positive muon ($\mu^+$) and an electron ($e^-$). Due to its purely leptonic composition, hadronic effects modify the atomic levels only as loop corrections and there are no finite-size effects, therefore it is an ideal system to test bound-state quantum electrodynamics (QED). Precision spectroscopy of the 1S-2S transition of M \cite{c0,c1} and the ground-state hyperfine splitting \cite{c2} contribute additionally to the most precise determinations of the electron-to-muon mass ratio and the muon magnetic moment. These measurements are also sensitive to the charge equality between muon and electron \cite{c1} and hence contribute to testing lepton universality, a topic that moved into the spotlight after the latest results of LHCb \cite{c2_11} and the Fermilab $g-2$ experiment \cite{c2_12}. Recently, precision measurements lead to new results on the ground-state hyperfine splitting in muonium measured at J-PARC~\cite{c2_14,c2_15} and on the Lamb shift in muonium measured at PSI~\cite{c2_17}. Further spectroscopy measurements using M atoms at improved levels of precision have been proposed \cite{c2_1,c2_1_1} and are currently being carried out \cite{c2_13,c2_16}. Another test of lepton universality is the search for the lepton-flavour violating muonium-anti-muonium oscillations \cite{c3}, for which a new measurement has recently been proposed \cite{c3_1}. At J-PARC, the ionization of M atoms in vacuum is part of a proposed cooling scheme for the $\mu^+$ beam \cite{c2_2}, which could enable next-generation searches for new physics. These experiments rely on high intensity and high quality atomic muonium beams in vacuum, and reaching higher precisions is partially limited by presently available vacuum M sources.\\
%
%
\begin{figure*}
	\centering
	\includegraphics[width=0.65\linewidth]{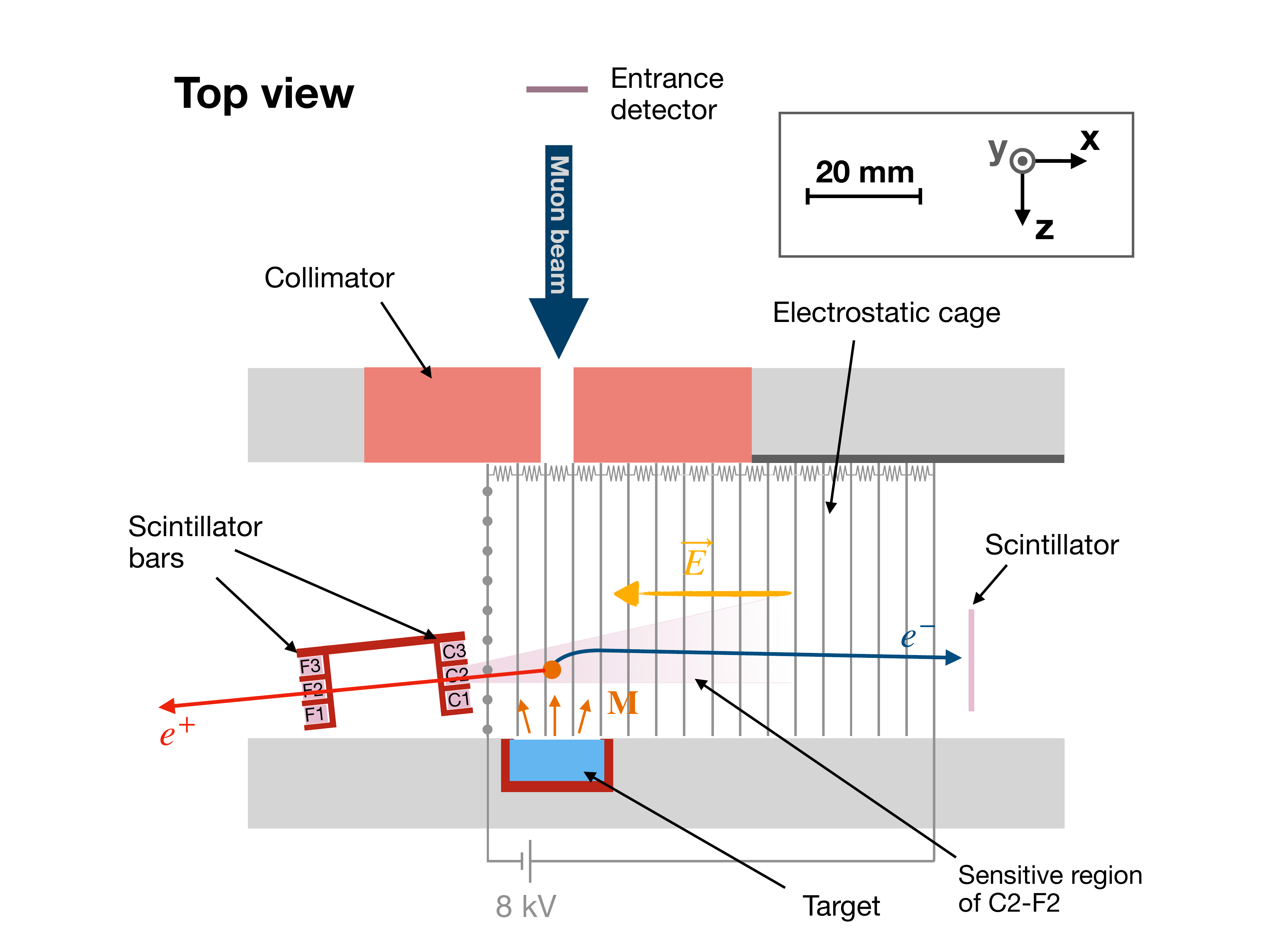}
	\caption{Top view of the experimental setup for the detection of M decay using scintillator bars and atomic electron detection. The muons entered through the entrance detector indicated at the top of the drawing. They formed M atoms in the target of which a fraction were emitted into vacuum. Decays of the M atoms in vacuum were detected by coincidence of the decay $e^+$ in the scintillator bars (red arrow) and atomic $e^-$ after electrostatic acceleration (blue arrow). The target was mounted on a PVC frame (grey box) outside of the electrostatic cage to repel ionization electrons.}
	\label{fig:expDrawingStabli}
\end{figure*} 
High yields of vacuum M are frequently produced by stopping low-momentum $\mu^+$ ($p_{\mu}< 28$~MeV/c) in porous materials like silica powders \cite{c5,c6,c7,c8} where they can combine with electrons. Here, the initial numbers of stopped muons per second ($\phi_{\mu}$) are converted with typically high efficiencies up to \SI{61}{\percent} \cite{c8_1}, but the flux of M atoms that can actually reach vacuum ($\phi_{\rm M_V}$) is strongly dependent on the diffusion times and the consequent decay losses due to the short muon lifetime ($\tau_{\mu} \approx 2.2~ \upmu$s). Hence the muon-to-vacuum-muonium conversion efficiency $\eta_{\rm M}=\phi_{\rm M_V}/\phi_{\mu}$ depends on the initial muon beam momentum that defines the implantation depth of $\mu^+$. Moreover, the temperature, chemical properties, and nanoscopic structure of the converter further impact the diffusion times. Note, that the stopping spread of muons within a certain material increases approximately proportional to $p_{\mu}^{3.5}$ below $p_{\mu} \approx 30$~MeV/c beam momentum~\cite{c8_2}. Consequently, lower beam momenta generally enable a higher fraction of the incoming muons to stop close to the sample's surface and be emitted into vacuum.
Recent developments for high M yields and beam qualities utilized mesoporous silica in cryogenic environments \cite{c9} as well as silica aerogels \cite{c10,c11}. High vacuum M rates could be achieved by laser ablation of microscopic holes in aerogels that decreased the M diffusion times, and reached conversion efficiencies of up to $\eta_{\rm M}\approx \SI{3.05(3)}{\percent}$~\cite{c12,c13}, using $p_{\mu} = 23~{\rm MeV/c}$ muons. One disadvantage of these sources is that the emerging M beam has a wide (approximately thermal Maxwell-Boltzmann) momentum distribution, and a large angular divergence $\propto \cos \uptheta$ measured relative to the surface normal. \\

%
%
Aiming for a muonium gravity experiment \cite{c14} and increased precision of M 1S-2S spectroscopy, the focus of our research started with the development of a novel cryogenic muonium source \cite{soter2021development}. The gravity experiment needs a compact detection setup and reliable background-free M detection methods that can ultimately be operated in a dilution refrigerator. As an alternative solution, a gravity measurement at higher temperatures may be possible using novel M emitters in combination with collimators. In order to study the characteristics of muonium emitters and test compact detection designs that can be adapted to a cryogenic environment, we developed various detection schemes and carried out measurements using known and novel room temperature M sources at the $\pi$E1 beamline of the Paul Scherrer Institute (PSI). \\
%
%
In Section~\ref{sec:scintillatorBars}, a compact background-suppressed M detection scheme is described which was used to characterize the dynamics of the emitted M atoms in vacuum. To our knowledge, our measurement was the first observation of M emission into vacuum from zeolite samples. The same setup was also used to quantify the impact of helium (He) gas on the mobility of emitted M atoms, which puts boundaries on the operational temperatures in our cryogenic experiments (Section~\ref{sec:scattering}). Furthermore, we carried out detailed emission studies using positron track reconstruction along one plane with MicroMegas tracker detectors~\cite{c14_1}, which allowed us to extract the muon-to-vacuum-muonium conversion efficiency for one aerogel target, discussed in Section~\ref{sec:MicroMegas}. 
%

\section{Background-free muonium detection}
\label{sec:scintillatorBars}
%
%
We developed a compact detection system based on tracking positrons ($e^+$) from $\mu^+$ decay, and a coincident detection of the low-energy atomic electrons left behind after M decay. The sketch of the experimental setup is shown in Figure~\ref{fig:expDrawingStabli}. Muons of $p_\mu = 11$-$13~$MeV/c momentum with a momentum spread of about \SI{8}{\percent} (full width at half maximum FWHM) from the $\pi$E1 beamline of PSI were guided to the experimental setup, where they first traversed a $55~\mu\text{m}$ thick scintillator foil (entrance detector) which was read out by a set of silicon photomultipliers (SiPMs). In this process the $\mu^+$ lost a large fraction of their kinetic energy and suffered significant scattering. Hence a passive copper collimator of 15~mm thickness with an opening of 6~mm width and 10~mm height was placed in front of the target area. The muons passing through the collimator reached the porous muonium conversion target with kinetic energies of few $100~$keV about 10~ns after the entrance signal, where they came to rest at different implantation depths below the surface, depending on their initial energy. Due to the distance between the entrance counter and the target, the rate of stopped muons on the target $\phi_{\mu}$ was not known, preventing the extraction of absolute conversion efficiencies $\eta_{\rm M}$. The stopped $\mu^+$ combined with electrons to form M atoms which could diffuse in the porous targets. The majority of M atoms decayed while diffusing through the sample, while a fraction reached the surface and were emitted into high vacuum, backwards relative to the muon beam, i.e. in $-z$ direction. The vacuum M atoms propagated with a certain velocity distribution, and passed in front of a scintillator tracker system, which consisted of small scintillator bars (3$\times$4$\times$25 mm$^3$) read out by SiPMs. The scintillators were arranged in two layers, next to the drift volume of the M atoms (C1-C3) and 25 mm further away (F1-F3), and detected the decay positrons from M (red arrow in Figure~\ref{fig:expDrawingStabli}). By requiring coincidences between the close and far scintillator bars (C and F) we defined three conical acceptance regions in the drift volume for muon decays (as indicated for C2-F2 by the transparent red cone in Figure~\ref{fig:expDrawingStabli}). The geometric acceptance for an isotropic decay in the center of these regions was roughly $\epsilon_{\rm geo} \approx 4\times 10^{-3}$. \\
The events considered as muonium emission candidates all started with an incoming muon triggering the entrance detector, followed by a coincidence measured in one of the scintillator pairs. The time difference between the entrance signal and the coincidence signal in the scintillator bars (typically up to $\sim \SI{8}{\upmu s}$) was used to determine the time it took for the M atoms to reach the acceptance region of the scintillator pairs. However, in addition to decays of vacuum muonium in the acceptance region, the coincidence scintillators registered a big muon-correlated background, i.e. a background decreasing exponentially with the muon lifetime.\\
\begin{figure*}
\centering
\includegraphics[width=0.95\linewidth]{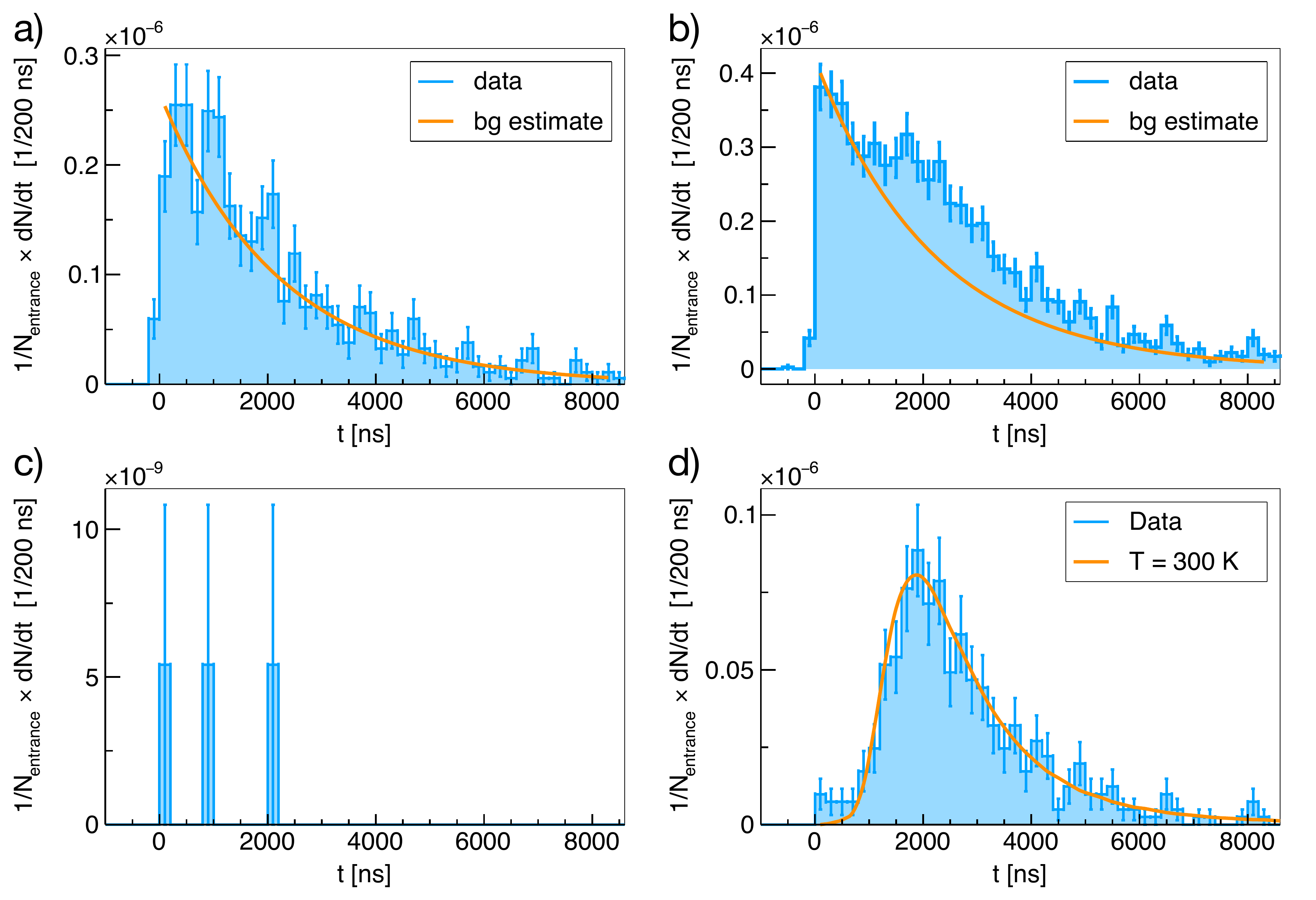}
\caption{Time distributions of $e^+$ signal in time coincidence normalized by the number of hits in the entrance detector. The given time is relative to the muon entrance. \textbf{Top:} Time distributions without additional coincidence with the atomic $e^-$ for the coincidence C2-F2 of the PVC measurement (a) and of the Aerogel-1 measurement (b). Here, the orange lines correspond to estimates of the exponential background from $\mu^+$ decay (see text for more details). \textbf{Bottom:} Time distributions including the additional coincidence to the atomic $e^-$ for the coincidence C2-F2 of the PVC measurement (c) and of the Aerogel-1 measurement (d). The orange curve in the distribution for aerogel corresponds to the simulated time distribution for M beam following a thermal, $\cos\uptheta$ emission at $300~$K.}
\label{fig:PVCAerogel}
\end{figure*}
In order to suppress this background, we required a coincidence with detection of the left-over atomic electron ($e^-$, blue arrow in Figure\ref{fig:expDrawingStabli}). A similar approach was already used to detect the decay electron in anti-muonium in~\cite{c3}. After dissociation of the M atom, the atomic $e^-$ have energies of a few eV, distributed around a mean energy of \SI{13.5}{\eV}~\cite{c14_2}. To create a measurable signal, we accelerated them towards a plastic scintillator plate with a surface area of $20\times \SI{20}{\square\mm}$ and a thickness of 1~mm using an electrostatic cage. Electron hits in the scintillator plate were registered by a Hamamatsu (R7600U) Photomultiplier tube (PMT). The electrostatic cage consisted of wires in front of the scintillator bars and a series of rectangularly shaped electrode loops surrounding the drift volume. The voltage of the wires in front of the scintillator bars was set to $-8~$kV and the voltage of the electrode loops decreased linearly towards ground on the side of the PMT. The cage was designed so as to repel ionization electrons that may have been emitted from the target with energies of several eV~\cite{c17}. In the following, the requirement of two scintillator bars triggering in coincidence with the atomic electron detector after a signal in the entrance detector will be referred to as the triple-coincidence requirement. \\
%
%
Using this setup, we investigated seven different M conversion targets. Two of these were laser ablated silica aerogel samples obtained by courtesy of G.~Marshall (TRIUMF) and co-workers. The 7~mm thick slabs of aerogel had average densities of 29~mg/cm$^3$ and featured microscopic vertical holes in the front surface. The holes were ablated with a laser in a triangular grid pattern~\cite{c12}. The holes of the Aerogel-1 sample were $4$-$5$~mm deep and had diameters of $100$-$110~\upmu$m and a pitch of 150~$\upmu$m. Aerogel-2 featured holes with diameters of $175$-$240~\upmu$m, a pitch of 550~$\upmu$m and an ablation depth of $\sim 1~$mm. Two M emitters were zeolite samples, which are composite materials of silicon, oxygen, hydrogen and aluminum featuring intrinsic microporosity with pores of about $0.5 $~nm size. The used zeolites were hierarchical zeolites which had been treated with alkaline to induce mesopores with $5$-$10~$nm size in the material. Zeolites are known positronium (Ps) production targets \cite{c17_2,Mitchell2018} and $\upmu$SR studies \cite{c15, c16} had already found that muonium atoms form in the bulk material when irradiated with $\mu^+$, but to our knowledge vacuum M emission has not been observed yet. The two samples used in this experiment were produced by P.~L.~Begona and co-workers at ETH Zurich and are referred to as Zeolite-1 (HZ40-AT2) and Zeolite-2 (CBV712-B) in this paper. They featured different nanoscopic structures, which were a MFI structure \cite{Kokotailo1978, Katzer1987} for Zeolite-1 and a FAU structure with cubic unit cells for Zeolite-2 \cite{Katzer1987}. We tested some more exotic samples, which were a carbon nanotube enforced silica aerogel, and two different carbon nanotube targets (single-walled CNTs, and CNT forests). No significant vacuum M emission was observed from any of these three samples. \\
The measurements further described here were conducted in two sets with similar beam conditions at $p_{\mu^+}\approx 11~\text{MeV/c}$ momentum. In the first set, the two zeolite samples were compared to Aerogel-1. In the second set, Aerogel-1 and Aerogel-2 were referenced to a PVC sample, that was assumed not to produce any vacuum M. Between the two sets the geometry was changed: An additional detector pair (coincidence C3-F3) was added to the setup and the target was moved closer to the scintillator bars (-$x$ direction) by 5 mm to optimize the tracking conditions.\\
Figure~\ref{fig:PVCAerogel} shows measured time distributions of the $e^+$ coincidences in the case of PVC (no M emission) and Aerogel-1, with and without the additional coincidence constraint with the atomic electron detector. The time on the horizontal axis refers to the time difference between the $\mu^+$ triggering the entrance detector and the decay $e^+$ being registered in the pair of scintillator bars. The time distributions of the $e^+$ coincidences feature a dominant exponential background, decaying with the lifetime of the muon. This background stems from decays of $\mu^+$ stopping in front of the scintillator bars, and from scattered decay $e^+$ coming from the target. Vacuum M atoms which were emitted from the target surface and pass by the sensitive region of the $e^+$ coincidence manifest as a bump on top of the exponential background. Using the additional coincidence with the accelerated atomic electron (triple-coincidence requirement) we obtain muonium distributions which are free of the muon-related background. Applying the triple-coincidence requirement to the PVC measurement (Figure~\ref{fig:PVCAerogel}c), only a few background hits remain. Using these hits we can define $\mathcal{P}^{BG}_{e^+e^-coinc}$ as the probability that a background hit in an $e^+$ coincidence is accompanied by a false muonium signal in the triple-coincidence distribution. It can be estimated by normalizing the background hits in the triple-coincidence distribution of PVC to the exponential background in the corresponding $e^+$ distribution. The resultant probabilities for the consecutive coincidence layers along the beam axis are given by
\begin{eqnarray*}
\mathcal{P}^{BG}_{e^+e^-coinc} = \left\{ \begin{array}{c} \left(3.7\pm1.8\right)\cdot 10^{-3}~~\left(\text{C1-F1}\right)\\ \left(6.5\pm 3.2\right)\cdot 10^{-3}~~\left(\text{C2-F2}\right) \\ \left(10.1\pm 4.1\right)\cdot 10^{-3}~~\left(\text{C3-F3}\right).\end{array}\right.
\label{eq:BGLevel}
\end{eqnarray*}
\begin{figure*}
\centering
\includegraphics[width=0.95\linewidth]{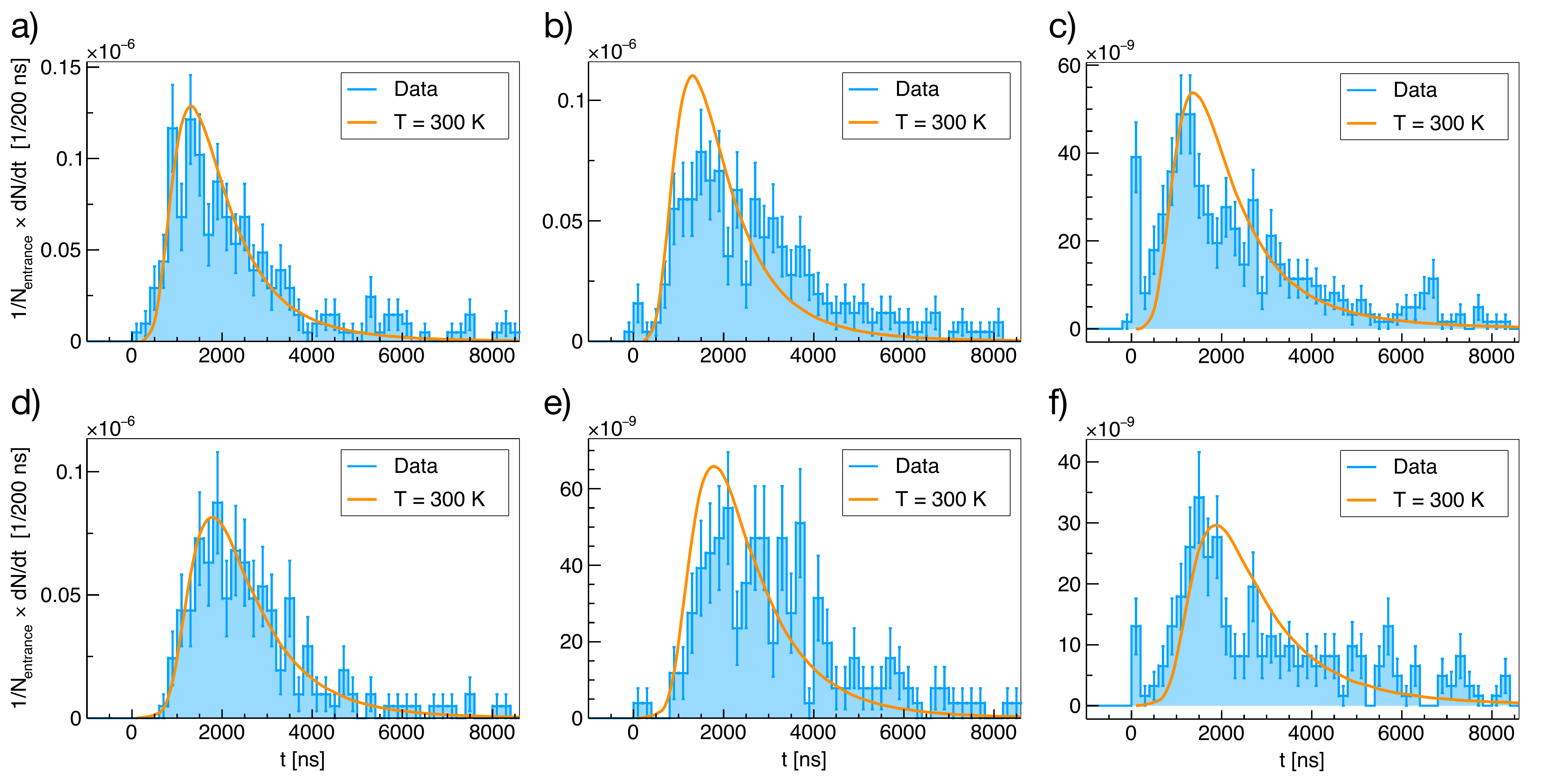}
\caption{Time distributions of the triple coincidence (positron in both scintillator bars and atomic electron detected) for C1-F1 (a-c) and C2-F2 (d-f). The measured data was obtained for Aerogel-1 of the second set (a,d), Aerogel-2 (b,e) and Zeolite-1 (c,f). The time is given relative to an entrance hit. The orange curve corresponds to the simulated time distribution for a M beam following a thermal $\cos\uptheta$ emission at $300~$K, with re-scaled amplitude to fit the data. }
\label{fig:ZEdis}
\end{figure*}
Based on Monte Carlo simulations, we suspect that the dominant source for the remaining background in the triple-coincidence distributions are not muon decays in the target, but muon decays occurring upstream in front of the collimator. This is a possible explanation for the increasing trend of $\mathcal{P}^{BG}_{e^+e^-coinc}$ with increasing distance from the target. In Figure~\ref{fig:PVCAerogel}d, a few remaining entries of the exponential background can be seen in the first bins after $t=0$. Similar background characteristics were observed for all coincidences of the aerogel measurements.\\
In order to study the dynamic properties of M emission we simulated the expected time distributions for M atoms emitted from the target. To obtain these time distributions and detection efficiencies, Monte Carlo simulations of the $e^+$ detector acceptances were combined with simulations of the atomic electron acceleration and simulations of the M atoms emitted from the sample. The electric field in the cage needed for the $e^-$ acceleration was simulated with a finite-element solver in COMSOL Multiphysics$^{\text{\textregistered}}$ \cite{c18}. The trajectories of atomic electrons were then simulated by tracking them through the fieldmap using GEANT4~\cite{c18_1} and G4beamline~\cite{c19}. We assumed the M atoms were emitted from the surface as a thermal beam, with a Maxwell-Boltzmann velocity distribution at 300~K, and corresponding angular distribution of the emission proportional to $\cos \uptheta\mathrm{d}\Omega$, where $\uptheta$ is the angle of the emission with respect to the target surface normal. In the simulations, the diffusion time of the M atoms inside the target was neglected ($\Delta t_\text{Diff}\approx 0$) compared to their time-of-flight (TOF) between the target surface and their decay position, in agreement with the observations. With this assumption, the starting time of the muonium atoms coincides with the time of the entrance signal, since the incoming muons only needed about 10~ns to travel from the entrance detector to the target sample. \\ 
The time distribution obtained from the simulations is superimposed on the triple-coincidence distribution of Aerogel-1 using an orange curve in Figure~\ref{fig:PVCAerogel}d with a re-scaled amplitude to best fit the data. The model is in good agreement with the measurement. Within the time interval $\left[600~\text{ns},6500~\text{ns}\right]$ the reduced $\chi^2$ is $\chi^2 / n_F \approx  36.2/30 \approx 1.21$. Here, the $\chi^2$ function in the limit of low counting statistics is used following \cite{c20}, which is given by
\begin{equation}
\chi^2=2\sum\limits_{i} N_i^{sim} -N_i^{exp} + N_i^{exp} \ln{\frac{N_i^{exp}}{ N_i^{sim}}},
\end{equation}
where $N_i^{sim}$ and $N_i^{exp}$ denote the numbers of entries in the i-th bin of the simulated and measured distributions respectively.
For other time distributions of Aerogel-1 at similar beam conditions we found reduced $\chi^2$ values between $\chi^2 / n_F = 0.7$ and $\chi^2 / n_F = 2.6$  (the lower interval limit has been adjusted to 500~ns, 600~ns and 1000~ns for the coincidences C1-F1, C2-F2 and C3-F3 respectively, to cut away remaining background counts at early times).\\ 
Further time distributions for Aerogel-1, Aerogel-2 and Zeolite-1 are displayed in Figure~\ref{fig:ZEdis}. In the distributions for Aerogel-2 the peaks are wider and occurring at later times. This indicates that the emission characteristics are different from those of Aerogel-1 and of the thermal emission model. Also the distributions for Zeolite-1 cannot be explained with a simple thermal emission and feature a faster component of muonium atoms that were not entirely thermalized. This can be understood qualitatively by the fact that the measured hierachical zeolites entail both micropores and mesopores. While in pores of 5~nm diameter M is expected and known to thermalize to the sample temperature \cite{c9}, for pores of 0.5~nm one needs to consider quantum mechanical effects since de Broglie wavelength of muonium is of the same order as the pore sizes. Therefore the lowest energy at which M can be emitted into vacuum is limited by the ground state energy of M in the pores. This behaviour is well studied for positronium which being approximately 100 times lighter already experiences this effect in mesoporous materials \cite{c20_1,c20_2}. Furthermore, we observed that all time spectra with zeolite samples feature a prominent prompt peak at $0<t<200~$ns. Note, that this peak has not been observed in any of the time spectra with aerogel samples. The peak might imply that a fraction of the muonium atoms are formed from muons backscattering at the surface of the zeolite sample while picking up an electron. \\
\begin{table}
\caption{\label{tab:yieldsStabli} Relative vacuum yields $\xi_{\rm M} = \eta_{\rm M}/\eta_{\rm M,0}$ for the various samples. The highest conversion efficiency $\eta_{\rm M,0}$ was achieved with the Aerogel-1 sample, which is thus used as a reference.}
\begin{ruledtabular}
\begin{tabular}{ccc}
& &   \\
\multicolumn{2}{c}{Set A} & $\xi_{\rm M}$   \\
\hline
Aerogel-1     & small holes     &  $ 1.0$ \\
Zeolite-1     & HZ40-AT2     &  $0.62\pm0.11$ \\
Zeolite-2     & CBV712-B     &   $0.66\pm0.12$  \\
\hline
&&\\
\multicolumn{2}{c}{Set B} & $\xi_{\rm M}$ \\
\hline
Aerogel-1     & small holes     &  $ 1.0$   \\
Aerogel-2     & large holes & $0.85\pm0.18$      \\
\end{tabular}
\end{ruledtabular}
\end{table}
The rate observed in the triple-coincidence distributions can be used to compare the conversion efficiencies $\eta_{\rm M}$ of the samples. As the number of muons stopping on the sample per second $\phi_{\mu}$ is not known precisely, we define the relative yield
\begin{equation}
    \xi_{\rm M}=\frac{\eta_{\rm M}}{\eta_{\rm M,0}},
\end{equation}
which relates the muon-to-vacuum-muonium conversion efficiency ($\eta_{\rm M}$) of a given sample to that of a reference sample ($\eta_{\rm M,0}$). Aerogel-1 featured the highest conversion efficiencies and is used as the reference sample. In order to keep the impact of different emission characteristics low we only consider the first coincidence layer (C1-F1) which is closest to the target surface to obtain $\xi_{\rm M}$. We define $R^{\text{M}}_{e^+e^-}$ as the number of detected triple coincidences normalized to the entrance counts. It is obtained by integrating the measured triple-coincidence distribution $H_{e^+e^-}\left(t\right)$ and subtracting the number of remaining background counts following
\begin{eqnarray}
&R^{\text{M}}_{e^+e^-} &= \int\limits_0^{t_{max}}H_{e^+e^-}\left(t\right)\mathrm{d}t\nonumber\\
&&-\mathcal{P}^{BG}_{e^+e^-coinc}\times\int\limits_0^{t_{max}}\text{bg}_{\text{e}^{+}}\left(t \right)\mathrm{d}t\text{,}
\label{eq:PVC_subtract}
\end{eqnarray}
where $t_{max}=8500~$ns was chosen as the upper integration limit. For this, the exponential background in the $e^+$ coincidence is estimated based on the number of hits in a small interval $\left[0,t_0\right]$ in the beginning of the time spectrum,
\begin{eqnarray}
\text{bg}_{\text{e}^{+}}\left(t \right) = e^{-\frac{t}{\tau_{\mu}}}\times \frac{\int\limits_0^{t_{0}}H_{e^{+}}\left(t^{\prime}\right)\mathrm{d}t^{\prime}}{\int\limits_0^{t_{0}}e^{-\frac{t^{\prime}}{\tau_{\mu}}}\mathrm{d}t^{\prime}}\text{,}
\label{eq:BGNorm}
\end{eqnarray}
where $H_{e^{+}}\left(t\right)$ is the measured time distribution of $e^+$ coincidences and it is assumed that the background is decaying with the muon lifetime $\tau_{\mu}=2197~$ns, as verified with the PVC sample. For the three rows of coincidences we set $t_0$ to 500~ns, 750~ns or 1000~ns respectively, such that in the interval $\left[0,t_0\right]$ the contribution of vacuum M decays is negligible. To correct for different dynamical behavior of the samples the results for $R^{\text{M}}_{e^+e^-}$ are additionally multiplied by a lifetime factor, such that the relative yield for any sample is calculated with
\begin{eqnarray}
\xi_{\rm M} = \frac{R^{\text{M}}_{e^+e^-}\times e^{\frac{\bar{t}}{\tau_{\mu}}}}{R^{\text{M,0}}_{e^+e^-}\times e^{\frac{\bar{t}_0}{\tau_{\mu}}}},
\label{eq:xi_formula}
\end{eqnarray}
where $\bar{t}$ corresponds to the mean of the time distribution measured with the sample and $R^{\text{M,0}}_{e^+e^-}$ and $\bar{t}_0$ indicate the values for Aerogel-1. The $\xi_{\rm M}$ obtained in this way are summarized in Table~\ref{tab:yieldsStabli}. The given uncertainties result from counting statistics. The $\xi_{\rm M}$ given for the zeolite samples correspond to an early measurement, which resulted in the highest measured $\xi_{\rm M}$ for these samples. Evaluating the yields of a later measurement, it was found that the relative yields of the zeolite samples were subject to a degrading effect and had dropped by up to $\sim \SI{30}{\percent}$ between the two measurements.\\
\begin{figure*}
\centering
\includegraphics[width=0.95\linewidth]{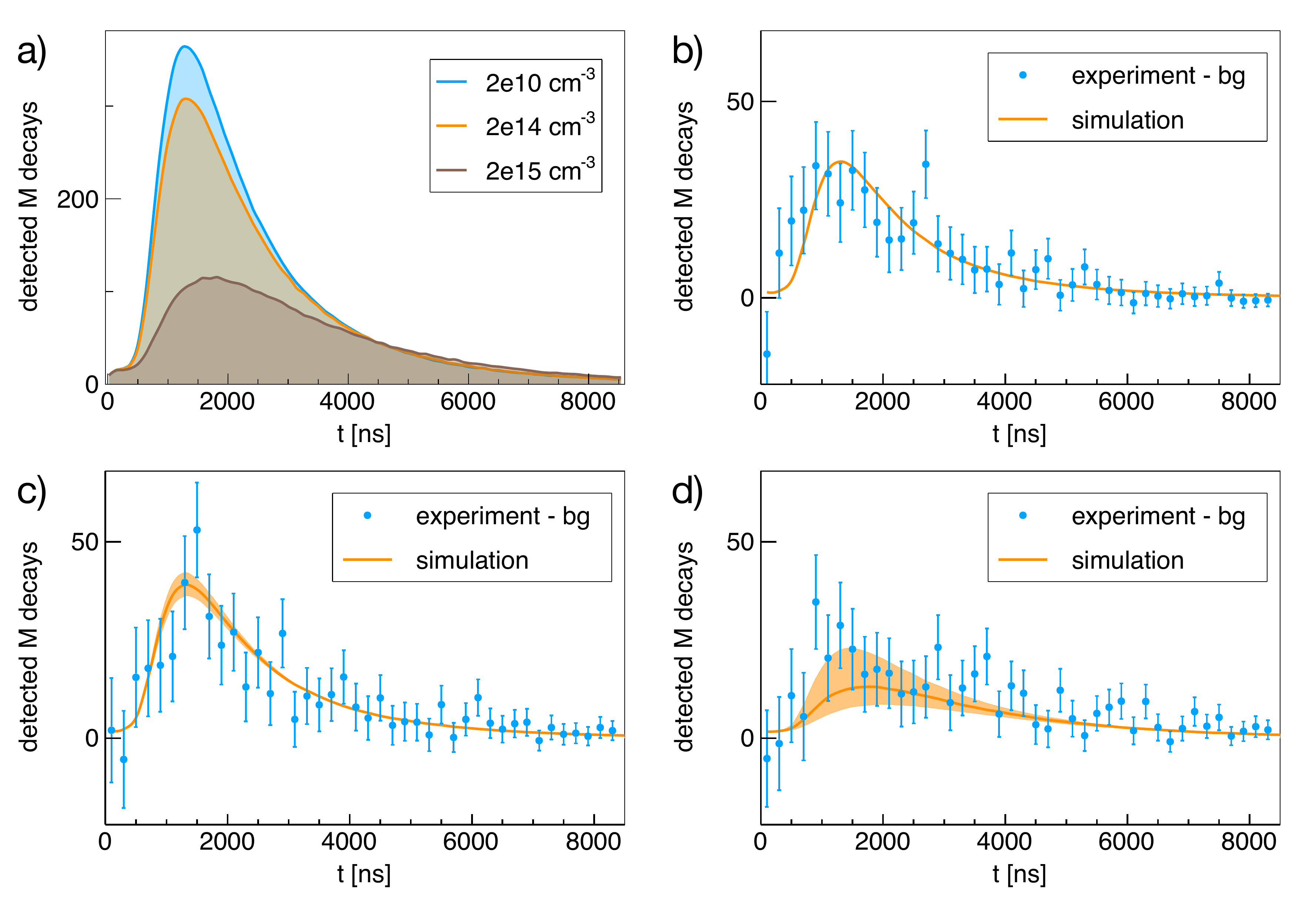}
\caption{Validation of the van der Waals elastic scattering model with measured time distributions of Aerogel-1. \\\textbf{a)} Time distributions for three He gas densities simulated for coincidence C1-F1 using the scattering model (arbitrary vertical scale). \textbf{b)-d)} Comparison of measured time distributions for coincidence C1-F1 with simulated distributions for the model at number densities b) $n_0 \approx 2\cdot 10^{10}~\text{cm}^{-3}$, c) $n_1 = \left(2\pm 1\right)\cdot 10^{14}~\text{cm}^{-3}$ and d) $n_2 = \left(2\pm 1\right)\cdot 10^{15}~\text{cm}^{-3}$. The measurement at $n_0$ is used to fix the global normalization of the simulation. The orange lines correspond to the central values for $n$ while the error bands account for the possible range given by the uncertainties of the absolute pressure determination.}
\label{fig:scatteringSpectra}
\end{figure*}
From a technical point of view, it is interesting to study the detection efficiency of the atomic electron detector. The overall efficiency of the detection system for the atomic electron $\epsilon_{e^-}$ consists of two individual efficiencies,
\begin{eqnarray}
\epsilon_{e^-}=\epsilon_{\rm geo}\times\epsilon_{e^- \rm det.},
\label{eq:efficiencyDef}
\end{eqnarray}
where $\epsilon_{\rm geo}$ is the geometric efficiency of an accelerated $e^-$ to reach the detector in case a positron coincidence was triggered, and $\epsilon_{e^- \rm det.}$ is the efficiency of the atomic electron detector (consisting of plastic scintillator and PMT) to detect the $\sim 6$-$8~$~keV electron. This overall efficiency can be assessed by comparing the number of $e^+$ coincidences triggered by M decays with the number of triple coincidences triggered by the same decays. In order to estimate the number of $e^+$ coincidences triggered by M decays, the $\mu^+$ background $\text{bg}_{\text{e}^{+}}\left(t \right) $ needs to be subtracted from the $e^+$ coincidences,
\begin{eqnarray}
R^{\text{M}}_{e^+} = \int\limits_0^{t_{max}}H_{e^+}\left(t\right)\mathrm{d}t-\int\limits_0^{t_{max}}\text{bg}_{\text{e}^{+}}\left(t \right)\mathrm{d}t\text{.}
\label{eq:subtractPosBG}
\end{eqnarray}
The overall efficiency of the atomic electron detection system can now be calculated for each coincidence pair via 
\begin{eqnarray}
\epsilon_{e^-}=\frac{R^{\text{M}}_{e^+e^-}}{R^{\text{M}}_{e^+}}.
\label{eq:totEff}
\end{eqnarray}
\begin{table}
\caption{\label{tab:efficienciesStabli}Measured and simulated efficiencies for the various rows of $e^+$ coincidence detectors. $\epsilon_{e^-}$ is the overall efficiency of the atomic electron detection system. $\epsilon_{\rm geo}$ is the efficiency of the acceleration of atomic electrons towards the detector.}
\begin{ruledtabular}
\begin{tabular}{ccc}
 & measured $\epsilon_{e^-}$  & simulated $\epsilon_{\rm geo}$  \\ 
 \hline
C1-F1 & $0.32 \pm 0.03^{\text{stat}}\pm 0.04^{\text{sys}}$ & $0.45\pm0.04$\\
C2-F2 & $0.34 \pm 0.03^{\text{stat}}\pm 0.01^{\text{sys}}$ & $0.50\pm0.02$ \\
C3-F3 & $0.33 \pm 0.04^{\text{stat}}\pm 0.01^{\text{sys}}$ & $0.43\pm 0.02$   \\
\end{tabular}
\end{ruledtabular}
\end{table}
Table~\ref{tab:efficienciesStabli} presents the weighted means over several measurements with aerogel targets. The systematic uncertainties emerge from the subtraction of the exponential background and were estimated based on the amount of entries within the interval $\left[0,t_0\right]$ of the triple-coincidence distributions.\\
By simulating the acceleration of the atomic electrons in the electric cage we can determine $\epsilon_{\rm geo}$ - which takes into account the coincident detection of a positron signal - and then extract the detection efficiency of the atomic electron detector $\epsilon_{e^- \rm det.}$. The computed geometric efficiencies $\epsilon_{\rm geo}$ for electron tracking are given in Table~\ref{tab:efficienciesStabli}. The given systematic uncertainties on the $\epsilon_{\rm geo}$ account for the uncertain number of positron coincidences following from early M decays, which depends on the applied energy cuts, the specific muonium emission model and the exact positioning of the detectors. Using the efficiencies of each coincidence row and computing the weighted mean, we find that the atomic electron detector detects the atomic electrons accelerated to kinetic energies of $6- 8$ keV with an efficiency of
\begin{eqnarray}
\epsilon_{e^- \rm det.} = \left(71 \pm 5\text{(stat)} \pm 6\text{(sys)}\right)\SI{}{\percent} .
\label{eq:PMTefficiency}
\end{eqnarray}

\section{Scattering of muonium atoms in helium gas}
\label{sec:scattering}
In a future muonium gravity experiment, residual helium gas may be present as vapor due to the use of superfluid He for M production \cite{soter2021development}. Therefore an estimate for the elastic scattering cross section of M atoms with the residual He gas will be needed in order to assess the risk of M atoms to scatter during the time of the free-fall measurement. For this estimation we used the same detection geometry described in Section~\ref{sec:scintillatorBars}. The measurement was performed by separating the vacuum of the target volume from the vacuum of the beam line, and applying He gas pressures of $p_1=\left(8\pm 4\right)\cdot 10^{-3}~$mbar, $p_2=\left(8\pm 4\right)\cdot 10^{-2}~$mbar as opposed to the high vacuum values ($p_0\approx1\cdot 10^{-6}~$mbar). The large uncertainties given here are not standard deviations but systematic uncertainties that reflect the accuracy of the absolute pressure determination for the pressure gauges used. The Aerogel-1 target was used for M production. Due to the He environment, the $e^+$ coincidences were used without the HV cage and atomic electron detector, which could not be operated with the buffer gas present.\\
We calculated cross sections for elastic M-He scattering assuming a simple spherical van der Waals interaction potential of purely electrical origin. The ionization energy, reduced mass and polarizability of the muonium system are approximately the same as in the hydrogen system and therefore the M-He potential is similar to the potential between atomic H and He, which has been studied extensively \cite{c21,c22,c23,c24,c25}. In the center-of-mass system (CMS), following \cite{c24}, we assumed the M-He potential to have a Lennard-Jones form,
  \begin{eqnarray}
V\left(r\right) = \varepsilon \left(\left(\frac{r_m}{r}\right)^{12}-2\left(\frac{r_m}{r}\right)^6\right),
\label{eq:LennardJones}
\end{eqnarray}
with $\varepsilon = 6.66\cdot 10^{-4}~$eV indicating a particularly shallow well and  $r_m = 3.46~\text{\AA}$. Solving the radial Schrödinger equation for this potential, we expanded the total cross section in CMS in partial waves using the expression
\begin{eqnarray}
\sigma_{tot}=\frac{4\pi}{k}\sum\limits_{l}\sigma_l\left(k\right)\text{,}
\label{eq:partialWave1}
\end{eqnarray}
where
\begin{eqnarray}
\sigma_{l}\left(k\right)=\left(2l+1\right)\sin^2\delta_l\left(k\right)\text{.}
\label{eq:partialWave2}
\end{eqnarray}
Here, $k$ is the wave number in CMS and $\delta_l$ is the scattering phase shift for the $l$th partial wave. Likewise, the differential cross sections were computed and transferred to scattering rates in the lab system at the corresponding temperature, which was done by averaging over the thermal momentum distribution of the He atoms. Since this model considers only two-body interactions it is valid for low gas densities at which the rate of 3-body interactions is negligible.  \\
In the simulations we assumed thermal M emission as described in Section~\ref{sec:scintillatorBars}. The simulations were run with He number densities $n_1 \approx \left(2\pm 1\right)\cdot 10^{14}~\text{cm}^{-3}$, $n_2\approx \left(2\pm 1\right)\cdot 10^{15}~\text{cm}^{-3}$ and $n_0 \approx 2\cdot 10^{10}~\text{cm}^{-3}$, roughly corresponding to the measured pressures at 300~K. During the vacuum measurement, the He pressure was low enough such that the uncertainty of the pressure measurement does not make a difference. Figure~\ref{fig:scatteringSpectra} shows time distributions for the coincidence C1-F1. In Figure~\ref{fig:scatteringSpectra}a, simulations for the three densities are compared. For the measured time spectra in the other panels we subtracted the exponential background which we determined as discussed in Section~\ref{sec:scintillatorBars}. The norm of the vacuum simulation was fitted to the vacuum measurement to extract a global normalization factor which was then used to normalize the simulations at number densities $n_1$ and $n_2$. The resulting curves are shown Figure~\ref{fig:scatteringSpectra}c)-d), where the orange error bands reflect the range given by the uncertainties of the absolute pressure determination. Considering the large statistical uncertainty of the measured points and the additional uncertainty of the background subtraction the computed cross sections reproduce the measurement with better than an order-of-magnitude level of accuracy. For a more precise test of the model, a dedicated high-statistics measurement at well-controlled densities would be necessary.\\
Using the computed cross sections we can estimate the maximally allowed density of residual helium in a gravity measurement with muonium. To ensure that the fraction of M atoms scattering on a length scale of about 100~mm (corresponding to the prospected length of the muonium free fall in the gravity experiment) is below $\sim$\SI{5}{\percent}, the helium number density needs to be reduced to around $10^{11}~\text{cm}^{-3}$. Hence, ignoring a possible 3He content, temperatures below $\approx 0.3~$K are needed to ensure accordingly low He saturated vapor pressures, as given by measurements and the Clausius–Clapeyron equation (see, e.g., Ref.~\cite{c26}). 

\section{Emission study with MicroMegas detectors}
\label{sec:MicroMegas}
\begin{figure}
	\centering
	\includegraphics[width=0.95\linewidth]{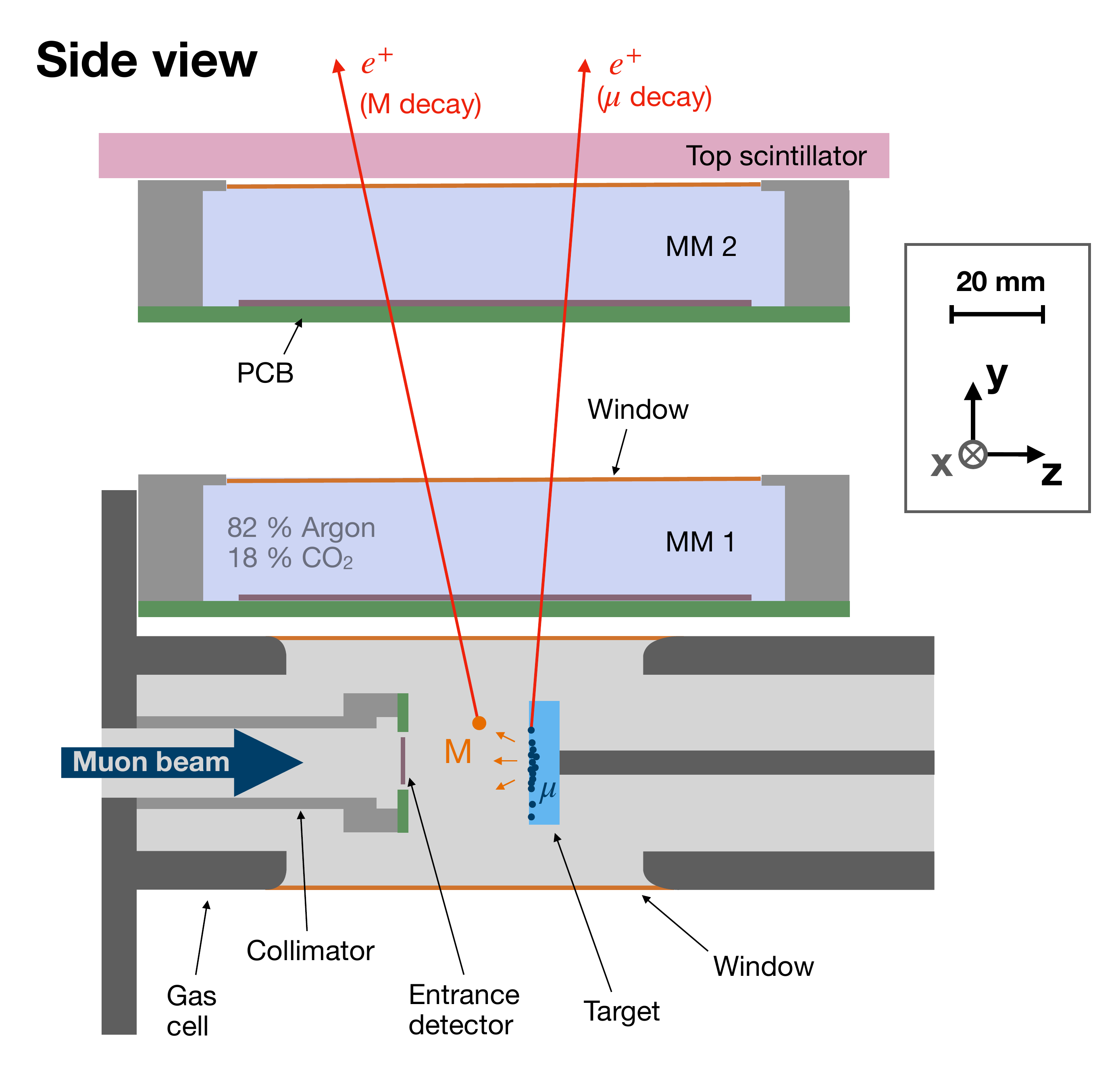}
	\caption{Experimental setup of the MicroMegas measurement. The $\mu^+$ entered from the left and triggered the entrance detector before reaching the target. Two MM detectors were mounted above the vacuum cell in a telescope configuration. A kapton window of \SI{130}{\upmu \m} thickness was glued onto the vacuum cell. A scintillation counter on top of the telescope was used to trigger the MM modules. For more details see text.}
	\label{fig:expDrawingMM}
\end{figure}

In our second experimental setup, a pair of MicroMegas (MM) tracking detectors \cite{c14_1} were used to study the muonium emission characteristics of aerogel more precisely. Previously, muonium emission measurements at TRIUMF \cite{c12,c13} used several layers of multi-wire drift chambers to track the decay positrons from muonium atoms. The MM modules we used had an active region of $80 \times$\SI{80}{\square\mm} and were developed at ETH Zurich for the NA64 collaboration~\cite{c27}. \\
A sketch of the experimental setup is shown in Figure~\ref{fig:expDrawingMM}. The muons passed through a collimator and triggered the entrance detector when entering the target cell. The aerogel target surface had a distance of $\sim 28$~mm from the entrance detector, which enabled around \SI{90}{\percent} of the muons exiting the entrance counter to stop on the target. The target sample was held in place by thin plastic parts from the bottom and the sides which had a minimal cross section seen from the front, minimizing the number of muons stopping on the holders to a negligible level. Target and collimator were mounted inside an aluminum cell that had an internal width and height of 70~mm and 40~mm, respectively. The windows on the top and bottom of the cell were kapton foils of $130$~$\upmu$m thickness that were glued onto the cell. The dimensions of the cell were chosen in a way, that the number of $\mu^+$ that stopped on the walls or windows in front of the target were negligible compared to the number of M decays in that region. Two MM modules mounted above the target cell were used as a telescope to track $e^+$ produced by $\mu^+$ and M decays inside the target cell. A large scintillator on top of the setup was used together with the entrance detector to trigger the MM detectors. The MM detectors were only triggered if the top scintillator received a hit in the time window $\left[1000,~2000\right]$~ns after the entrance detector. In that way, we selected events in which muonium atoms had enough time to form, leave the sample and travel a few millimeters in vacuum before decaying. The data acquisition system of the MM detectors could record signals only within a limited time window of 675~ns around the trigger, and was limited to a trigger rate of 1~kHz. Due to these limitations in the data readout, the triggering scheme was crucial in order to separate M decays in front of the target from $\mu^+$ decays on the target surface.\\
\begin{figure}
\centering
\includegraphics[width=0.95\linewidth]{fig_6.pdf}
\caption{Distribution of $\mu^+$ and M decays projected to the central plane of the vacuum cell. $e^+$ tracks with angles of less than $\ang{10}$ to the vertical are considered and extrapolated towards the central plane. \textbf{a)} Projected 2D map of $\mu^+$ and M decays. The two accumulations of hits correspond to decays in the entrance detector and the target surface. \textbf{b)} Projection of all decays within $-10\leq x\leq 10~$mm. The black histogram corresponds to the measured data while the orange curve corresponds to combined simulations of background and M emission at $300~$K. The dashed blue line corresponds to the background simulation (without vacuum M signal). The blue error band accounts for the systematic uncertainty of the background simulation, which was estimated using the discrepancy between measured and simulated distribution on the right side of the target peak.}
\label{fig:MM_2D1D}
\end{figure}

\begin{figure*}
\centering
\includegraphics[width=0.95\linewidth]{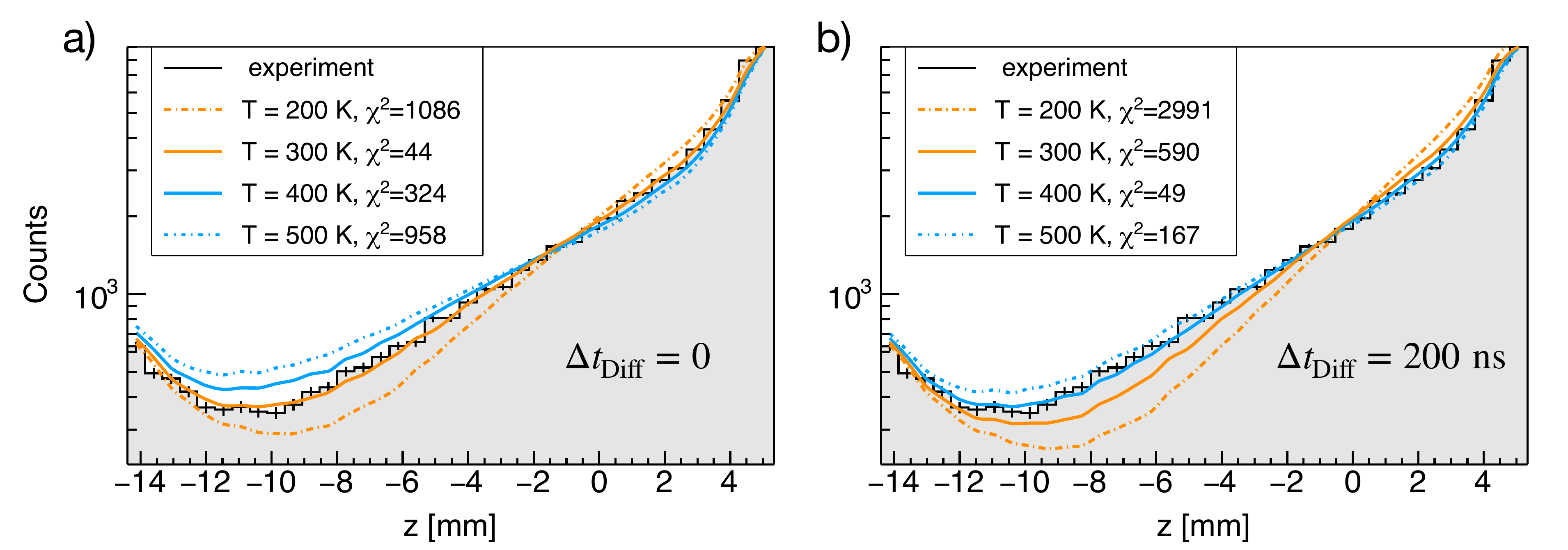}
\caption{Fits of emission models with different temperatures to the data measured with the MicroMegas detectors. \textbf{a)} Models without any diffusion time in the sample, i.e. $\Delta t_{\text{Diff}}=0$. \textbf{b)} Same models but with a constant diffusion time of $\Delta t_{\text{Diff}}=200~$ns added to the time of travel of the M atoms. The given $\chi^2$ values correspond to the fits within $\left[-12,~4\right]$~mm, with $n_\text{df}=29$ degrees of freedom.}
\label{fig:MM_1D_zoom}
\end{figure*}
\begin{figure}
	\centering
	\includegraphics[width=0.99\linewidth]{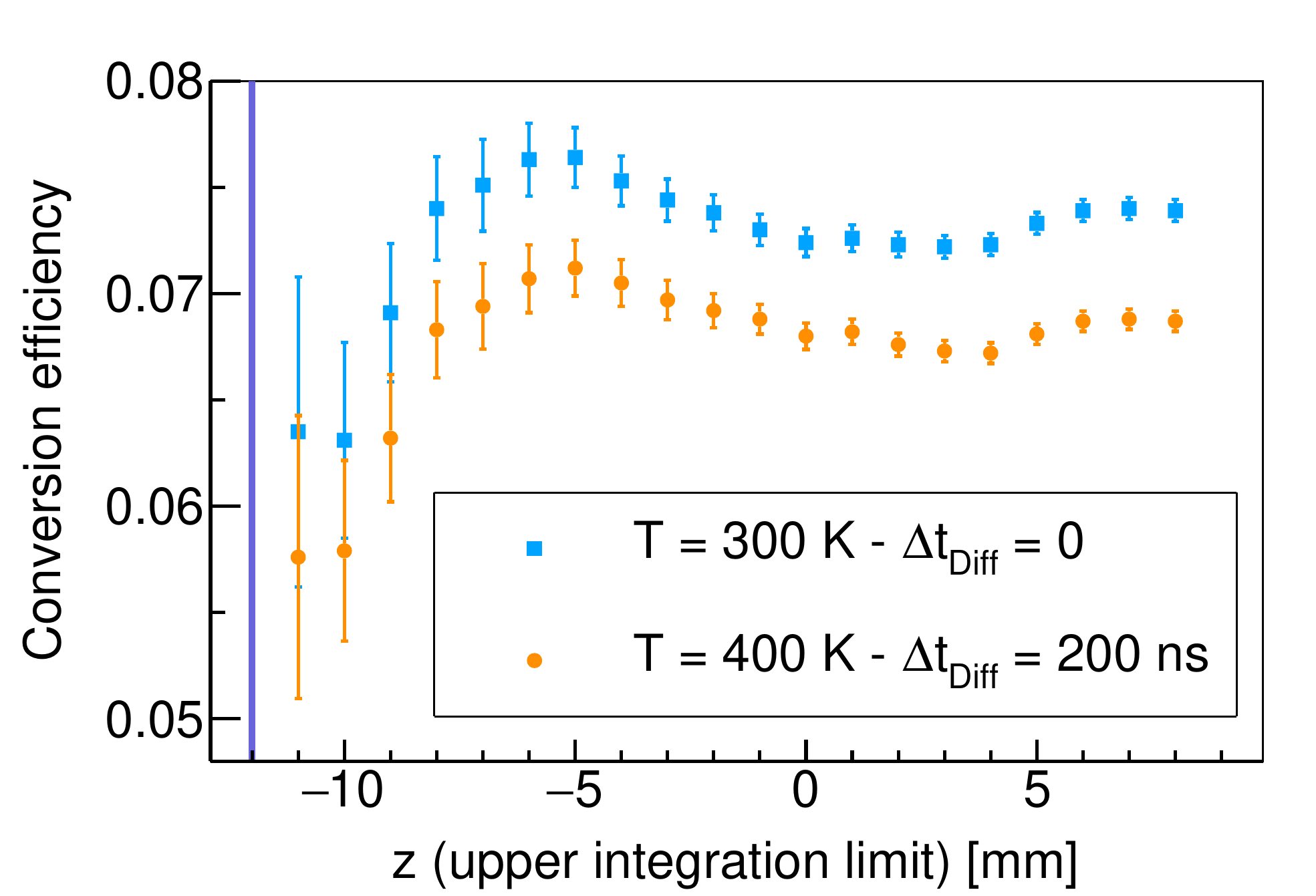}
	\caption{Dependence of model-dependent conversion efficiencies on the upper limit of the fit range. The lower limit is fixed at $z=-12~$mm. The two models with the best fits are compared. The error bars contain the statistical uncertainty only.}
	\label{fig:MM_yield_iB}
\end{figure}

The MM telescope allowed to produce an image of decays projected to the central plane of the target cell. To restrict the perspectival discrepancy between projected and actual decay position to less than 2~mm for decays in front of the target, we considered only tracks within 10 degrees of the vertical. The projected hit map of decays is shown in Figure~\ref{fig:MM_2D1D}a) with a logarithmic color scale. The large hit accumulation on the right corresponds to decays of stopped $\mu^+$ and M on the target surface, while the left accumulation corresponds to $\mu^+$ decays in the entrance detector. Left of the target peak, the cloud of M decays in vacuum is visible. \\
We again used Monte Carlo simulations to study the emission characteristics of the aerogel sample. For the background simulations we considered $\mu^+$ decaying at the target surface and the entrance detector as well as $\mu^+$ that missed the target and stopped on the walls or the windows. In the simulations, the thickness of the PCB in the MM modules was adjusted, such that the width of the simulated peaks matched the measurement. For unknown reasons, the PCB in the simulation needed to be thinner by roughly a factor of two than those actually mounted in the MM detectors, which were around $3.4$~mm thick. Figure~\ref{fig:MM_2D1D}b) shows a projection of the measured data onto the $z$ axis (black binned data). The dashed blue line between the peaks shows the distribution obtained from the normalized background simulations. The light blue error band indicates the systematic uncertainty of the background level in this region, which was estimated conservatively based on the discrepancies between the background simulation and the measurement on the right side of the target peak.\\
An excess of the measured data over the background simulation between the two peaks indicates the cloud of M atoms decaying in front of the target. Using the integral of the excess we can estimate a model-independent vacuum yield for the emission of muonium from the aerogel sample. For this, we subtract the integrals of the simulated background distribution $N_{\text{BG}}$ from the experimental distribution $N_{\text{exp}}$ to obtain the number of detected muonium decays within an interval $z\in \left[-12,~4\right]$~mm,
\begin{eqnarray}
N^{\text{M decays}}_{\left[-12,~4 \right]} = \int \limits_{-12}^{4}\left(N_{\text{exp}}\left(z\right)-N_{\text{BG}}\left(z\right)\right)~\mathrm{d}z\text{,}
\end{eqnarray}
where the interval was chosen such that the entries within the interval are dominated by muonium decays. Normalizing this number to the total number of decays on the target and in the cloud we obtain the vacuum yield for Aerogel-1 emitting M into $z\in \left[-12,~4\right]$~mm within the time window $t\in\left[1000,~2000\right]$~ns:
\begin{eqnarray}
\text{yield}_{\left[-12,4 \right]} =\frac{N^{\text{M decays}}_{\left[-12,4 \right]}}{N^{\text{M decays}}_{\left[-12,4 \right]}+N_{\text{target}}}\text{,}
\end{eqnarray}
where $N_{\text{target}}$ is the integral of the background distribution of simulated muon decays on the target. With this procedure we obtain 
\begin{eqnarray}
\text{yield}_{\left[-12,4 \right]} = \left(4.57 \pm 0.07\text{(stat)}~ ^{+0.63} _{-0.39}\text{(sys)}\right)\SI{}{\percent}\text{,}
\end{eqnarray}
where the uncertainty is dominated by the systematic uncertainty which we assigned to the background level between the two peaks (indicated by the light blue error band in Figure~\ref{fig:MM_2D1D}b)). \\
In addition to the model-independent approach we performed simulations of the M cloud assuming M emission at various temperatures and angular distributions. Fitting the amplitude of the obtained distributions we can determine which emission model fits best. The orange graph in Figure~\ref{fig:MM_2D1D} corresponds to the best fit result which yields a reduced $\chi^2$ of $\chi^2 / n_\text{df}\approx 44 / 29$ in the fit range $\left[-12,~4\right]$~mm. For this fit, emission of thermal M at $300$~K with a $\cos\uptheta$ distribution was assumed and the diffusion time of the M atoms inside the sample was neglected. Additionally, fits for emission models with other temperatures were performed. For higher temperatures, fits could be obtained by assuming non-zero diffusion times, which demonstrates correlation between the emission temperature and the diffusion time. It turned out that an increase in the diffusion time of about 200~ns was needed to compensate for a temperature increase of 100~K. Figure~\ref{fig:MM_1D_zoom} shows fitted distributions for various temperatures with an additional constant diffusion time of $\Delta t_{\text{Diff}}=200~$ns. The reduced $\chi^2$ values for the best fits at the corresponding temperatures are summarized in Table~\ref{tab:yieldsMM}. Above 400~K, the fit quality was found to decrease drastically, which implies that M atoms were emitted with roughly thermal energies and after short diffusion times of maximally a few 100~ns. In order to determine the diffusion time in the sample and the temperature quantitatively, it is necessary to vary the time window of the measurement. This could not be done during this study due to the limited run time and the constraints of the MM DAQ. Since the measurement was performed using a hardware trigger, it was also not possible to vary the timing after the measurement. For future measurements of this kind a variation of the time window is certainly needed. In general, the best fits were found for models with emission following a $\cos\uptheta$ angular distribution.\\
From the fit of the simulated distributions to the data, it is possible to extract a model-dependent conversion efficiency for the sample which we obtain as
\begin{eqnarray}
\eta_{\rm M} = \frac{N^{\text{vacuum}}_{\text{M, 0}}}{N^{\text{vacuum}}_{\text{M, 0}}+N^{\text{target}}_{\mu\text{, 0}}},
\end{eqnarray}
where $N^{\text{vacuum}}_{\text{M, 0}}$ is the absolute number of M atoms emitted into vacuum from the sample and $N^{\text{target}}_{\mu\text{, 0}}$ is the number of stopped muons remaining on the target. Both numbers can be extracted from the simulations normalized according to the fit result.\\
\begin{table*}
\caption{\label{tab:yieldsMM} Reduced $\chi^2$ and model-dependent conversion efficiencies $\eta_{\rm M}$ of Aerogel-1 for M emission models with various temperatures. The conversion efficiencies were obtained by fitting the simulated M distributions (assuming thermal M emission at $T=300~$K with a $\cos\uptheta$ distribution) to the data measured with MicroMegas detectors. The beam momentum in the measurement was at $p\approx 12.5~$MeV/c.}
\begin{ruledtabular}
\begin{tabular}{cccc}
		Temperature~[K]    &   Diffusion time~[ns]  &   $\chi^2~/ ~n_\text{df}$    &   $\eta_{\rm M}$~[\%]           \\
\hline \\
300     &  0   & $44~/~29$  & $ 7.23 \pm 0.05 \text{(stat)} ^{+1.06}_{-0.76}\text{(sys)}$   \\
400     & 200  & $49~/~29$   & $ 6.72 \pm 0.05 \text{(stat)} ^{+1.06}_{-0.76}\text{(sys)}$   \\
500     & 400  &  $184~/~29$   & $ 6.38 \pm 0.05 \text{(stat)} ^{+1.58}_{-1.29}\text{(sys)}$   \\
\end{tabular}
\end{ruledtabular}
\end{table*}
The conversion efficiency obtained in this way depends on the applied fit range. While the lower limit of the fit was fixed to $z=-12$~mm, which is where the M signal becomes negligible compared to the tail of decays from the entrance detector, the upper limit for the fit was varied to study the impact on the conversion efficiency. Figure \ref{fig:MM_yield_iB} shows the dependence of the obtained conversion efficiencies on the upper limit of the fit for the model with $300~$K and no diffusion time and the model with $400~$K and a constant diffusion time of $200~$ns. The latter model is accompanied by slightly lower conversion efficiencies due to the losses during the time of the diffusion. In order to extract the conversion efficiency, the fit interval $\left[-12,~4\right]$~mm was chosen as a benchmark, which contains the region in which the M distribution is dominant. The uncertainty due to the choice of the integration limit is taken into account as a systematic uncertainty which is estimated with the range of conversion efficiencies between the fit ranges $\left[-12,~4\right]$~mm and $\left[-12,~-3\right]$~mm (smaller fit ranges cut away most of the M distribution). Another systematic uncertainty is given by the discrepancy between the background simulation and the measurement which was discussed above and is indicated by the blue band in Figure \ref{fig:MM_2D1D}.
The conversion efficiencies obtained in this way for models with various temperatures are summarized in Table~\ref{tab:yieldsMM}. The latest publication of measurements at TRIUMF~\cite{c13} reports conversion efficiencies of about 1-\SI{2}{\percent} for comparable laser-ablated aerogels with similar hole arrangements. Their best conversion efficiencies, achieved with a sample with larger holes, was reported as \SI{3.05(3)}{\percent}. In comparison, the higher conversion efficiencies reported here (in Table~\ref{tab:yieldsMM}) benefit from the narrower stopping distribution of muons in the sample due to the low muon momentum of around $p\approx 12.5~$MeV/c (momentum spread $\sim \SI{8}{\percent}$ FWHM).

\section{Summary and conclusion}
We developed two compact detection systems to characterize the propagation of M atoms in vacuum, and compared various known and novel vacuum muonium emitters. The emission of M atoms from state-of-the-art laser-ablated aerogel samples was found to be in agreement with roughly thermal emission and the expected $\cos\uptheta$ angular distribution. These samples reached the highest muon-to-vacuum muonium conversion efficiencies in our studies. Zeolites that were previously optimized for positronium conversion were found to emit M atoms into vacuum as well, with somewhat non-thermal energy distributions. These samples provided a \textit{relative} vacuum M yield of $66\pm \SI{12}{\percent}$ with reference to the highest-yield aerogel sample, but their performance was not stable over time. \\
For the first time we used a pair of MicroMegas detectors to study emission of muonium into vacuum. This preliminary experiment allowed us to extract an absolute conversion efficiency for an aerogel sample and, more in general, served as a testing ground for future muonium experiments. 
Based on this experience, the MicroMegas telescope and the developed analysis tools have been integrated into the setup which is presently aiming at the measurement of the 1S-2S transition of M~\cite{c2_1} to monitor M production. \\

\noindent We developed detailed Monte Carlo simulations to compare the measured data with thermal emission models. In order to be able to describe the scattering of M atoms in residual He gas in a future cryogenic gravity measurement we calculated elastic M-He scattering cross sections based on a simple Lennard-Jones model. The theoretical cross sections were consistent with our measurements of M emission, where different amounts of He gas were introduced in the vacuum chamber. This measurement placed an upper bound of \SI{0.3}{\K} to the maximum temperature allowable for supporting a gravity measurement with muonium produced from a superfluid helium source.

\begin{acknowledgments}


The authors would like to thank the HIPA facility at PSI for the stable beam and technical support. Special thanks to the PSI detector group, M.~Hildebrandt, A.~Stoykov and F.~Barchetti, and U.~Greuter from the electronics group. We are indebted to S.~Kamal, G.~Marshall, T.~Mibe and co-workers for providing us the laser-ablated aerogel samples, to P.~L.~Begona at ETHZ for the zeolite samples, moreover to M.~de~Volder from Cambridge for the CNT samples. We would like to thank G.~Janka for providing support with the MicroMegas dectectors. The work of L.~Gerchow was supported by the ERC consolidator grant 818053-Mu-MASS.

\end{acknowledgments}

\bibliographystyle{unsrt}

\end{document}